# Scaling the α-relaxation time of supercooled fragile organic liquids

C. Dreyfus[1], A. Le Grand[1], J. Gapinski[2], W. Steffen[3] and A. Patkowski[2]

[1] P.M.C., U.M.R.7602, Case Postale 77, UPMC, 4 Place Jussieu, 75005 Paris, France.

[2] Institute of Physics, A. Mickiewicz University, Umultowska 85, 61-614, Poznan, Poland.

[3] Max Planck Institute for Polymer Research, Ackermannweg 10, 55128 Mainz, Germany.

Abstract :

It was shown recently [1] that the structural α-relaxation time τ of supercooled *o*-terphenyl depends on a single control parameter Γ, which is the product of a function of density $E(\rho)$, by the inverse temperature $T^{-1}$. We extend this finding to other fragile glassforming liquids using light-scattering data. Available experimental results do not allow to discriminate between several analytical forms of the function $E(\rho)$, the scaling arising from the separation of density and temperature in Γ. We also propose a simple form for τ(Γ), which depends only on three material-dependent parameters, reproducing relaxation times over 12 orders of magnitude.

I-Introduction

The steady increase over more than 12 orders of magnitude of the shear viscosity η, or of the structural α-relaxation time τ measured by dielectric or light-scattering spectroscopic methods, is among the spectacular features that accompany the liquid-glass transition [2]. The latter can take place in a large variety of liquids ranging from mineral oxides, salts, organic polymers, metallic alloys, to simple molecular liquids. Under cooling, these liquids avoid crystallisation at the melting temperature $T_m$ and become a glass at a lower temperature $T_g$. For low molecular weight glassforming materials, a glass is usually defined as a liquid whose relaxation times τ are larger than $10^2$ seconds, a phenomenon that takes place at some temperature $T_g$. In spite of its frequent occurrence and of continuous theoretical and experimental efforts over the last fifty years, this transition remains one of the least understood phenomena in condensed matter physics. In particular, one still does not fully grasp how and on which parameters, usually called control parameters, these relaxation times τ depend [3, 4]. Obviously, in simple glassforming colloids the only control parameter is the density. Conversely, would the structural relaxation of a liquid at equilibrium be only due to jumps over energy barriers that are independent from temperature and density, τ might have a simple Arrhenius behaviour and the only control parameter would be temperature. The Angell plot [2], which represents the logarithm of relaxation times τ (or viscosity η) as a function of $T_g/T$, shows at ambient pressure a continuous change from strong glasses (in the Angell classification) such as $SiO_2$, where the Arrhenius behaviour is approximately met, to fragile glasses, such as molecular liquids, where a strong positive curvature appears. From such non Arrhenius plots, one could infer the existence of two control parameters: density, ρ, and temperature, $T$, expressed as $T_g/T$. Nevertheless, there is an interesting *a priori* possible intermediate case which interpolates between the two preceding ones. There, $\tau$ would be expressed as a function of $\Gamma$ where $\Gamma$ is a product of $T^{-1}$ by a function of density. For dimensional reasons we write $\Gamma=E(\rho)/T$, where $E(\rho)$ is an energy depending only on the density. Within this framework the problem splits into three questions : (i) does such a scaling with a single control parameter $\Gamma$ exist? If yes, (ii) what is the expression for $E(\rho)$ and (iii) how does the dependence $\tau(\Gamma)$ look like ?

For the time being, there is no theoretical answer to these questions but recent experimental data can be analyzed along these lines. Indeed, new experimental developments have allowed to perform viscosity [5, 6] as well as light-scattering [7-15] and dielectric measurements [16-19] at variable temperature and pressure. Thus, the exploration of a large portion of the phase diagram and the measurement of the corresponding quantities is now possible. The present paper aims at partly answering these questions. Other recent works[1, 20-24] have already dealt with the same issue. Some of us have shown [1] that the relaxation times in molecular liquid OTP probed over 12 orders of magnitude by means of light-scattering experiments depend on a single control parameter $\Gamma$ of the form $\Gamma=\rho^n/T$ with $n=4$. This result extends a scaling property obtained on the same system by Tölle et al. [20] using an inelastic incoherent neutron-scattering technique and spanning a smaller time range. Two recent papers [21, 22] have extended and generalized the OTP result : analyzing dielectric relaxation (DR) times obtained either on polymers or on van der Waals molecular liquids extending over 7 to 10 orders of magnitude, they showed that the data can be scaled using control parameters of the same form as in OTP, $n$ becoming material dependent. Previously, a scaling was shown to hold in glycerol by Alba-Simionesco et al [23]. Also ref. [24] showed that the domain of density explored in the experiments was too small to ascertain the analytical form of $E(\rho)$. It is thus clear that a scaling property might exist at least for a large class of molecular liquids and for some polymers so that one can now concentrate on the two next questions. The present paper extends the previous studies [21, 22] in two different directions. In Part II, using dynamic light-scattering (DLS) as used for OTP, we look for the scaling properties. We confirm the results of ref. [22], obtaining scalings of the form $\rho^n/T$ where $n$ is material dependent and show that the values of $n$ are identical by the two spectroscopic methods (DR and DLS). Second, we explore the issue of the analytical form of $E(\rho)$ and we show that the available experimental results do not allow to discriminate between different forms. Part III deals with the more difficult third question to which we give a preliminary experimental answer. We propose a form of $\tau(\Gamma)$ which depends explicitly on two material parameters and implicitly on a third one, $\Gamma_g$, the value of the control parameter at the glass transition. We check the validity of this form over 12 orders of magnitude for several systems and show that, when used for ambient pressure data, it gives a better description of the Angell plot than the usual Vogel-Fulcher-Tamman (VFT) law. A conclusion offering perspectives for future work ends the paper.

II- Dependence of the α-relaxation on a control parameter

II-1 Scaling with a power law in density

Recent experiments [1] on OTP have shown that the α-relaxation times obtained by DLS over a very large time range ($10^{-10}$ s - $10^2$ s) could be scaled using a single parameter $\Gamma=\rho^4/T$. A generalization $\Gamma=\rho^n/T$ has since been proposed by Tarjus et al. [24] and it gives a scaling for the dielectric data on polymers [21]. DLS Photon-Correlation Spectroscopy (PCS) experiments give access to a narrower range of relaxation times ($10^{-5}$-$10^2$s), i. e. they probe only the strongly supercooled part of the phase diagram. This reduced time range might be insufficient to ascertain the value of $n$. Figure 1a shows that this is not the case : limiting the OTP data to the PCS results [7] gives a value of $n$ between 4 and 4.2, the uncertainty being not larger than for the whole time domain [1]. As the form $\Gamma=\rho^n/T$ with adjustable $n$ offers a large flexibility in the dependence of $\Gamma$ on density, this observation allows us to look for similar scalings of PCS data in other molecular fragile glassformers. Indeed, PCS relaxation times at various temperatures and pressures and Pressure-Volume-Temperature diagrams have been recently published for three glassforming liquids: diglycidil ether bisphenol A (DGEBA) [8], phenolphtaleine-dimethyl ether (PDE) [12, 13], 1,1 di ( 4-methoxy-5-methylphenyl) cyclohexane (BMMPC) [14]. The glass transition temperatures and their variation with pressure are listed in Table 1. At each pressure and temperature, the corresponding density can be obtained from the Tait interpolation formula, whose form and parameters are also collected in Table 1.

As exemplified in [1], representing the α-relaxation time as a function of ρ at constant temperature results in a set of distinct curves. When the relaxation times of these glassforming liquids are plotted as a function of $\Gamma=\rho^4/T$ (Figure 2, a-c), *i.e.* with the *n* value found in OTP, the different α-relaxation times almost collapse on a master curve for PDE and DGEBA but not for BMMPC. However, if *n* is allowed to vary (Figure 3, a-c), for each liquid a value of *n* can be found that gives an excellent scaling. The values of *n* for these liquids, based on literature data, are reported in Table 2 as well as that obtained from unpublished PCS data on BMPC (1,1-bis(p-methoxyphenyl)cyclohexane ) [25]. We add also the *n* value obtained by the scaling of the dielectric α-relaxation time in KDE (cresolphtaleine-dimethyl ether) [26]. The values of *n* vary from 3.8 to 7.5.

We note that similar results were obtained independently by Casalini and Roland [22] on DR data. Decoupling between DR and DLS experiments has sometimes been reported [10]. It does not seem to be the case here. Such a decoupling is also absent in the case of OTP : scaling of DR data in the same time range (Figure 1b) yields a slightly different value of *n* (4.25 instead of 4<n<4.2), but this difference is within experimental accuracy. We also checked for a decoupling of relaxation times from viscosity [6] (Figure 1c), where deviations from the expected behavior were noted in ref. [24]. Yet, these discrepancies occur in the pressure range where a polynomial approximation of the equation of state [27] was used for extrapolation, *i.e.* beyond the experimental range of the PVT measurements (the viscosities are measured up to 1 GPa while the equation of state is only measured up to 100 MPa). Using the Tait equation with the parameters reported in Table 1 in the domain inside which the PVT data were obtained, then calculating the density from this Tait equation results in a rather satisfactory scaling of the viscosity with $\Gamma=\rho^4/T$ (Figure 1d). We conclude that a scaling obtained from low temperature in van der Waals glassforming liquids can be extrapolated to higher temperature, and that light-scattering, dielectric, and viscosity data yield the same scaling, provided that a reliable equation of state is available.

II-2 Other scalings

In Section II-1 we have shown that the parameter $\Gamma=\rho^n/T$ leads to a scaling of the α-relaxation times over a broad time range for several glassformers, the value of $n$ being material dependent. This result is not specific to van der Waals liquids of nearly spherical molecules: DGEBA is an oligomer with a rather long chain. Also, as seen in Figure 3d, the viscosity data reported by Cook et al. in glycerol [5] scale with $n=1.8$, a value much smaller than found in other liquids. Glycerol is controlled by the dynamics of hydrogen bonds. Both examples make questionable the link between $n$ and the results obtained in soft-core systems [28] which have been at the origin of the educated guess made in refs. [20] and [1].

Tarjus et al. [24] proposed that a scaling is obtained if parameter $\Gamma$ is simply the product of a function of the density by the inverse temperature. They called the numerator of $\Gamma$ an "effective interaction energy, $E(\rho)$", and argued that the variation of density explored in the experimental range is too small to discriminate between different analytical expressions of $E(\rho)$. Following Figure 8 of ref. [24], we show in Figure 4 the function $E_1(\rho)\sim\rho^4$ over a density range similar to that spanned in experiment [1], as well as a linear function $E_2(\rho) \sim (\rho/b -1)$ with $b$ chosen to approximate $E_1(\rho)$ in the region of interest. Such a function emphasizes the role of a state with a low density equal to $b$ where $E_2(\rho) = 0$ while another $E_3(\rho) \sim \rho_0\rho/(\rho_0-\rho)$, based on a free volume model, would introduce the existence of $\rho_o$, a particular high density value. One sees in Figure 4 that parameters $b$ and $\rho_o$ can be found such that the three functions are very close. Figures 5 and 6 show that the relaxation times in BMMPC and PDE scale equivalently well with $E_1(\rho)$, $E_2(\rho)$ or $E_3(\rho)$. The values of the parameters $b$ and $\rho_0$ are reported for all these liquids in Table 2 together with the range of densities explored in the corresponding experiments.

II-3 Scaling of the liquid glass transition temperature $T_g$

The variation with pressure of the liquid-glass transition temperature $T_g$ could possibly lead to a more precise determination of the analytical form of $\Gamma$. Customarily, $T_g$ is defined as the temperature at which the relaxation time $\tau=100$ s. This corresponds to a constant $\Gamma_g=E(\rho_g)/T_g$ $T_g(P)$ can be calculated from $T_g$ and $\rho_g$ at ambient pressure, plus the equation of state [8, 13, 18] and one of the parameters $n$, $b$ or $\rho_0$ entering into the effective interaction energy considered.

In figure 7 we plot $T_g(P)$ calculated for PDE with the three effective interaction energy functions against the experimental $T_g$. The pressure $P$ varies from 0.1 MPa to 200 MPa, which corresponds to a total shift of $T_g$ by about 50 °C. The three functions yield closely similar liquid-glass transition temperatures, all presenting very similar deviations from the experimental values at 200 MPa. Here as well, the available data do not allow to discriminate between the three functions from the shift of $T_g$. We note that the experimental data show systematic changes of $\Gamma_g$ by about 2% over the entire pressure range. In this paper we assume that $\Gamma_g$ is simply constant and this leads to the deviation from the diagonal seen in Figure 7.

II-4 Conclusion of Part II

Neither the accuracy of the scaling law itself, nor the variation of $T_g$ with pressure allow us to distinguish between different analytical expressions for $\Gamma$. This does not mean that the particular scaling function proposed in [1] and generalized in [24] should be definitely discarded. In the very few cases where a direct numerical estimate of $E(\rho)$ is known [24], several liquids (all of them van der Waals liquids) seem to show a positive curvature of their effective energy with increasing density reminiscent of the $E_1(\rho)$ variation. In particular, values of E($\rho$) obtained by these authors in OTP agree perfectly with function $E_1(\rho)$ after normalization at $\rho_g$=1.145g cm$^{-3}$, as seen in Fig. 4. However, the simple factorization between density and temperature in Γ leads to much more general consequences than could be expected if the control parameter was only related to soft core systems properties [1, 20, 28]. If the only requirement is the existence of a positive mean slope of $E(\rho)$ around an operating density, the density variation in $E(\rho)$ can as well arise from other, more chemical origins : for example increasing density may increase the barriers to rotation in hydrogen-bonding systems or increase the barriers between different conformations in polymers; in both cases, this would lead to an increase of the effective interaction energy. This may explain why a single control parameter Γ also exists in other glassforming liquids, such as glycerol or, as recently shown, in polymers [21].

Finally, the α-relaxation time τ is not the only variable characterizing the α-part of the structural relaxation. The stretching parameter β is another important variable. The weak variation of β with pressure and the scattering of the experimental points in the few available studies (see for example refs. [12, 17, 19]), do not rule out a scaling of β with Γ.

III - Analytical expressions of the dependence of the α-relaxation times on density and temperature

III-1 Analytical Ansatz

The thermal variation of viscosity or relaxation times at constant pressure is usually expressed by the Vogel-Fulcher-Tamman law (VFT) or by the Williams-Landel-Ferry law although other expressions can be found [29]. All these expressions yield good fits to the relaxation-time or to viscosity data over limited temperature ranges but they usually suffer from systematic deviations from experiments over broader ranges, as was shown for example in ref. [29] through different representations of the data.

We know from Part II that $\tau$ depends on the single control parameter $\Gamma$. This should appear explicitly in the analytical expression. This requirement is not met in a transparent way in the laws mentioned above but it is fulfilled by the expression given in refs. [23, 24]. Following these authors, we derive a general expression for the relaxation time consistent with all experimental findings. First, one can always write the relaxation time as a function of temperature and density in the form :

$$\log(\tau/\tau_0) = \mathcal{E}(\rho, T)/T \qquad (1)$$

where $\tau_o$ is a constant. Eq(1) introduces the notion of an activated process where equilibrium is reached through jumps over energy barriers $\mathcal{E}(\rho, T)$ which depend on density and pressure (here log is $\log_{10}$). Second, all relaxation times are found to follow approximately an Arrhenius law at high temperature [29-32]. An expression of Eq. (1) which fulfills these two requirements and agrees with the results of part II is :

$$\log(\tau/\tau_0) = [E(\rho)/T\,]\Phi(E(\rho)/T) \qquad (2)$$

where $E(\rho)$ is defined in refs. [23, 24] as an energy associated with the barriers to α-relaxation and viscous flow in the high temperature regime, *i.e.* where the thermal variation of the relaxation time is Arrhenius like. $\tau_o$ is the corresponding relaxation time, typically of the order of 0.01 ns, while $\Phi(\Gamma)$ tends to 1 when $\Gamma$ tends to zero (*i.e.* for $T>>T_g$ $\Phi(\Gamma)$ is a positive monotonously increasing function of 1/T as known from the Angell plot. These two conditions are met by postulating $\Phi(\Gamma)= \exp(B\Gamma)$, where $B$ is a material-dependent positive constant. Then :

$$\log(\frac{\tau}{\tau_o}) = \frac{E(\rho)}{T}\exp(B\frac{E(\rho)}{T}) = \Gamma\exp(B\Gamma)\,; \qquad (3)$$

This expression contains only two system-dependent parameters $\tau_o$ and $B$, while the other system-dependent parameters (for example $n$, $b$, or $\rho_0$) are included into $E(\rho)$, as discussed in Part II. The behaviour at high temperature (in the liquid phase) and at low temperature (around $T_g$), easily extracted from Eq. (3), is discussed below.

III-2 Comparison with experimental results.

A convincing check of Eq. (3) must take into account the behaviour of the α-relaxation time at high and low temperature as well as the characteristic curvature observed in fragile liquids (usually between $10^{-8}$s and $10^{-6}$s). We consider first the OTP data obtained at variable temperature and ambient pressure and, second, the dielectric measurements [29] performed at ambient pressure in the liquids considered in Part II.

To this effect, it is convenient to introduce the reduced parameter:

$$\tilde{\Gamma} = \frac{\Gamma}{\Gamma_g} = \frac{E(\rho)}{T} \frac{T_g}{E(\rho_g)} \qquad (4)$$

where $\Gamma_g$ is the value of $\Gamma$ corresponding to the relaxation time $\tau_g$ at the liquid-glass transition. Then Equation (1) becomes:

$$\log(\tau/\tau_0) = \Gamma_g \tilde{\Gamma} \exp(B\tilde{\Gamma}\Gamma_g) \qquad (5a)$$

At the liquid-glass transition, $\log(\tau_g/\tau_0) = \Gamma_g \exp(B\Gamma_g)$, so that:

$$\log\left(\frac{\tau}{\tau_0}\right) = \log\left(\frac{\tau_g}{\tau_0}\right) \tilde{\Gamma} \exp\left(\tilde{B}(\tilde{\Gamma}-1)\right) \qquad (5b)$$

where $\tilde{B} = B\Gamma_g$ and where $\tilde{\Gamma}$ increases towards 1 when $T$ decreases from values typical of the liquid to the liquid-glass transition. Defining the glass transition temperature as the temperature for which the relaxation time is 100 s, we obtain $\log(\tau_g/\tau_0) \sim 13$ for typical cases. We also note that the $\tau_g$ dependence in Eq. (5b) is only apparent. Changing $\tau_g$ to another value only introduces another value of $\Gamma_g$, but does not modify the form of the equation. To test the validity of Eq. (5b), we need to introduce an analytical expression for $E(\rho)$. In the following we choose to take $E(\rho)=E_1(\rho)\sim\rho^n$, as already used in several papers [1, 21, 22, 24]. This yields:

$$\log\left(\frac{\tau}{\tau_0}\right) = \log\left(\frac{\tau_g}{\tau_0}\right)\frac{T_g}{T}\frac{\rho^n(T)}{\rho_g^n(T)} \exp\left(-\tilde{B}\left(1 - \frac{T_g}{T}\frac{\rho^n(T)}{\rho_g^n(T)}\right)\right) \qquad (6)$$

III-2-1 Comparison with light-scattering OTP data

In Figure 8, we show the light scattering data of OTP at ambient pressure [33] together with the curve obtained with $\tilde{B}=3.48$ and $n=4$ corresponding to the best fit to the experimental data. The data span more than 12 orders of magnitude and cover the high and low temperature regimes as well as the temperature domain where the Angell plot exhibits the very strong curvature typical of fragile liquids. We illustrate the respective roles of $\tilde{B}$ and $n$ in Eq. (6) by showing also in Fig. 8a and b the variation of $\log(\tau/\tau_0)$ with $T/T_g$, first keeping $n=4$ constant and letting $\tilde{B}$ vary, then keeping $\tilde{B}=3.48$ constant and letting $n$ vary. Both parameters give similar effects although not of the same magnitude. When n varies from 0 to 7.5 at constant $\tilde{B}$, the relaxation time at $T_g/T=0.85$ decreases by only ~ 4 orders of magnitude, while it decreases by ~ 6 orders of magnitude when $\tilde{B}$ varies from 0 to 3.48 (highest value found here) at constant $n$. Moreover, as both parameters produce similar effects, a variation of $n$ can easily be counterbalanced by a variation of $B$. It is only through variable density experiments that the two parameters could be properly determined.

III-2-2 Comparison with ambient pressure dielectric data

Let us now use Eq. (6) to fit dielectric α–relaxation times obtained at ambient pressure. This requires the knowledge of both the equation of state of the system and the value of $n$. This is possible for glycerol, PDE, BMPC, BMMPC and KDE for which we have determined $n$ and dielectric relaxation times [32] at ambient pressure are known over a very broad range of relaxation times. In the fit, the three[*] fitting parameters are $T_g$, $\tau_o$, and $\tilde{B}$. The fits are shown in Fig. 9, they correspond in each case to variations of τ over ~ 12 orders of magnitude. The values of $T_g$, $\tau_o$ and $\tilde{B}$ are reported in Table 3. $\tau_o$ is always of the same order of magnitude as relaxation times in normal liquids. Inside the error bar, the value of $T_g$ is consistent with experimental values (Table 1). Table 3 contains also the parameters of the fit of the data with a VFT law, $\log\tau/\tau_{VFT}=A/(T-T_o)$, also shown in Fig. 9. As usual, $\tau_{VFT}$ take unphysically small values, contrary to $\tau_o$, and $T_o$ is located well below the liquid-glass temperature and cannot be attained experimentally. Also, in all cases, the mean square deviation (m.s.d.) is smaller in the fits with Equation (6) than with a VFT law, with the same number of adjustable parameters. Equation (6) reproduces better the experimental results than the VFT law and its parameters have a clearer physical meaning.

III-2-3 The Stickel's representation of $\tau$

Stickel *et al*. [29b] proposed to discuss the systematic deviations from the experimental data presented by the empirical expressions of the $\alpha$-relaxation time (or viscosity) by plotting $\left(\partial\log\tau/\partial T\right)^{-1/2}$ versus $T$. This function was designed to emphasize these discrepancies and, in particular, to compare accurately experimental results with the VFT law whose representation is a straight line that passes through zero at $T_o$. As this formula involves the temperature derivative of the experimental relaxation time, it requires high-quality data. It is a strong test for any empirical law usually designed to approximate an unknown function, but not to fit also its first derivative. Introducing Equation (6) yields:

$$\left(\frac{\partial\log\tau}{\partial T}\right)^{-1/2} = \left[\log\left(\frac{\tau_g}{\tau_0}\right)\left(1+\tilde{B}\right)\left(T^{-1}+n\alpha\right)\tilde{\Gamma}\ \exp\left(\tilde{B}(\tilde{\Gamma}-1)\right)\right]^{-1/2} \quad (7)$$

where here $\alpha$ is the thermal expansion coefficient at constant pressure. The right-hand side of Eq. (7) is plotted in Fig. 10 for the same liquids as in Fig. 9, using the parameters reported in Table 3. Although the agreement with experimental data points, in particular in the region where they deviate from the VFT approximation, is only qualitative, the plot of Eq. (7)

---

* $T_g$ is a third fit parameter, as it enters into the value of $\Gamma_g$ or, rather, $\Gamma/\Gamma_g$

exhibits the same trends as the experimental points. This shows essentially that with only three adjustable parameters it is possible to find a function sufficiently flexible to exhibit trends observed in the experimental data that are not reproduced by the VFT law. One can infer from this result that the deviation from VFT observed in the experimental data is presumably related to specific characteristics of the structural relaxation time itself and not to experimental inaccuracies.

III-3 Limiting behaviour of $\tau$

III-3-1 High temperature

The Arrhenius like behavior of the α-relaxation time at high temperature is well documented both numerically and experimentally. For example, in a recent simulation of a mixture of van der Waals molecules [34], a deviation from Arrhenius behaviour was observed at decreasing temperature and constant density. Also an Arrhenius behaviour of the relaxation time as a function of pressure and temperature has been observed in the lowest viscosity (highest temperature) range of many glassforming liquids [29-32].

By construction this behaviour is contained in Eq. (6). At high temperature and constant density, $\rho_{liq}$, the expansion of Eq. (6) yields:

$$\log\left(\frac{\tau}{\tau_0}\right) = \log\left(\frac{\tau_g}{\tau_0}\right)\frac{A_{liq}}{T}\left(1 + \frac{\tilde{B}A_{liq}}{T}\right)\exp(-\tilde{B}) \quad (8)$$

where $A_{liq} = E(\rho_{liq})T_g/E(\rho_g)$ is a constant. The leading term is linear in $1/T$ with a positive $(1/T)^2$ correction. At constant pressure, as $\rho$ is an increasing function of $T_g/T$, further positive contributions in $(T_g/T)^2$ appear: deviations from the Arrhenius behaviour towards a positive curvature are more pronounced in plotting constant-pressure results than constant-density results. However, the variation of the density is weak enough for the linear behaviour to be observed over a reasonable temperature range.

III-3-2 Fragility

Eq. (6) was designed to reproduce the upward curvature of the relaxation times on the Arrhenius plot observed in fragile liquids, which leads to a large effective activation energy close to $T_g$. The activation energy at constant pressure is usually characterized by the fragility coefficient $m$ [35] :

$$m = \left( \frac{\partial \log \tau}{\partial (T_g / T)} \right)_{T=T_g} \quad (10)$$

Introducing Eq. (5b) yields:

$$m = \log\left(\frac{\tau_g}{\tau_0}\right)\left(1 + \tilde{B}\left(1 - \frac{T_g}{E(\rho_g)}\left(\frac{\partial E(\rho)}{\partial T}\right)_{T=Tg}\right)\right) \quad (11)$$

Taking $E(\rho) = E_1(\rho)$, one can compute the fragility $m$ from the values of $\log(\tau_g / \tau_0)$, $n$, $\tilde{B}$, $T_g$ and the thermal expansion coefficient $\alpha$ :

$$m = \log(\tau_g / \tau_0)(1 + \tilde{B})(1 + T_g n \alpha) \quad (12)$$

The values computed at ambient pressure are reported in Table 2, and agree satisfactorily with the experimental values. An activation curve measured at constant density would give a different, density-independent, fragility coefficient $m_d = \log(\tau_g / \tau_0)(1 + \tilde{B})$, always smaller than $m$, a result already obtained in ref. [21, 24].

IV – Conclusion

We have shown that a scaling of relaxation times measured in light scattering at variable temperature and pressure can be obtained over a very broad range of times with a single control parameter. This confirms other recent studies. This control parameter is the product of the inverse of temperature by a function of density, whose analytical form cannot yet be made precise. Beginning with a first general formulation, we also proposed an analytical form for the relaxation time as a function of the control parameter which includes by construction its variation with density. This expression is shown to describe the variable temperature and pressure data over 12 orders of magnitudes of time provided that the equation of state is known. Without any further measurement, it is able to give a good estimate of the relaxation time as a function of temperature and density.

Several aspects of the results presented here call for further developments. Experimentally, it will be important to investigate the reliability of the proposed scaling at higher pressure. The few available viscosity measurements suggest that this scaling could be valid up to several GPa, but confirmation from spectroscopic data is obviously needed while, so far, these techniques are not available in this pressure range. Another issue is a more precise determination of the proper analytical form of the activation energy $E(\rho)$, in particular in van der Waals liquids. Eventually, this might establish a differentiation between several classes of glasses through this function. More generally, the validity of this scaling in strong glassforming liquids such as oxide glasses is an open question. Although the experimental requirements are difficult to meet, the answers to these questions seem important enough to be an incentive for developing high pressure techniques.

Acknowledgment: C.D. and A.L.G warmly acknowledge in-depth discussion with R. Pick on the material contained in this paper as well as for stimulating suggestions and criticisms and are very grateful to C. Bousquet for her constant help. C.D. thanks H.Z. Cummins for many discussions during her visit in New York. C.D and W.S. also thank E.W. Fischer for advices and discussions. Partial financial support of the Committee for Scientific Research, Poland (KBN, Grant No. 1 P03B 083 26) is gratefully acknowledged. Collaboration within the three groups was made possible by the bilateral agreement between CNRS and Max-Planck Society and by the Polonium cooperation agreement no 075546UE between France and Poland.

Table captions

Table 1

Liquid-glass transition temperature, density and their derivatives, and coefficients of the Tait equation.

The dependence of the reciprocal density $V$ on pressure $P$ is approximated by the Tait equation[36]:

$$V(T,P) = V(T,0)(1 - 0.0894\ln(1+P/B(T)))+V_{\text{ref}},$$

with $V(T,0) = \nu_0 + \nu_1(T-273.15) + \nu_2(T-273.15)^2$ and $B(T) = b_0\exp(-b_1(T-273.15))$, $T>273$ K. The parameters $\nu_0$, $\nu_1$, $\nu_2$, $b_0$ and $b_1$ of the Tait equation are given in the Table. For BMMPC : $V_{\text{ref.}}=0.21364$ cm$^3$/g and for BMPC, $V_{\text{ref.}}=0.192585$ cm$^3$/g [37] where $V_{\text{ref}}$ is a correction recommended by the authors of ref 18 [37]. $V_{\text{ref}} = 0$ in the other cases. Values for d$\rho$/d$T$ and d$^2\rho$/d$T^2$ are computed from $V(T,P)$.

References are a) [18], b) [14], c) [17], d) [41], e) [26], f) [42], g) [12], h) [13], i) [40], j) [8,9,10], k) [44], l) [5].

Table 2

Parameters $n$, $b$ and $\rho_o$ appearing in $E_1(\rho) \approx \rho^n T^1$, $E_2(\rho) \approx \rho/b-1$ and $E_3(\rho) \approx \rho\rho_0/(\rho_0-\rho)$. and density range spanned in the experiments. Parameter $\tilde{B}$ of Equation (6), fragility $m$ calculated from Eq. (11) and experimental fragility $m$ obtained by DLS in all cases except KDE (where $m$ is obtained from DR) and glycerol (where $m$ is obtained from viscosity or DR). References are a) [14], b) [26], c) [42], d) [12], e) [43], f) [8], g) [5,39].

Table 3

Fits of experimental α-relaxation times [32] taken at ambient pressure with Eq. (6) (first four lines) and VFT law (last four lines) : values of the fitting parameters and of mean square deviations ( m.s.d.).

Figure captions

Figure 1: Scaling of the relaxation times and viscosity in OTP.

a) relaxation times from photon-correlation spectroscopy [7].

b) relaxation times from dielectric spectroscopy [16].

c) viscosity obtained from the Naoki polynomial approximation (see text) [27].

d) viscosity obtained from the Tait equation (see text).

Figure 2: Scaling of α-relaxation time in different glassformers with $\rho^4/T$.

a) PCS data in PDE[12] : 34.2°C : ¤ ,44.8°C: o, 54.6°C : △, 65.3°C : ▽, 75.2°C : ◇

b) PCS data in DGEBA[8] : 0.1 MPa : ¤ , 20 MPa : o, 40 MPa :△, 60 MPa: ▽, 80 MPa: ◇, 100MPa : ◁, 120 MPa : ▷, 140 MPa : ⌂, 160 MPa : ☆.

c) PCS data in BMMPC[14] : 0.1 MPa : ¤ , 20 MPa : O, : 40 MPa △, 60 MPa: ▽, 80 MPa : ◇, 100 MPa : ◁, 120 MPa : ▷, 140 MPa : ⌂, 160 MPa : ☆, 180 MPa : ⌂, 200M Pa : O.

d) viscosity data in glycerol[5] : 0°C : ¤ , 12.5°C : O, 22.5°C : △, 50°C : ▽, 75°C : ◇

(Here to be compared to Figure 3d).

Figure 3: Scaling of the same data as in Fig. 2 with $\rho^n/T$. The value of $n$ is indicated for each material.

Figure 4: Comparison between the three energies curves $E_1(\rho) \approx \rho^4$ (———), $E_2(\rho) \approx \rho/b - 1$ (······), and $E_3(\rho) \approx \rho\rho_0/(\rho_0-\rho)$ (-------) ( $E_1(\rho)$ is normalized at 1 at $\rho_g = 1.145$ g cm$^{-3}$, $E_2(\rho)$ and $E_3(\rho)$ are fitted to $E_1(\rho)$). Direct estimate of $E(\rho)$ from ref. [24] : O.

Figure 5: BMMPC : scaling with a) $E_2(\rho) \approx \rho/b - 1$ and b) $E_3(\rho) \approx \rho\rho_0/(\rho_0-\rho)$. Same data as in Figure 2c.

Figure 6: PDE : scaling with a) $E_2(\rho) \approx \rho/b - 1$ and b) $E_3(\rho) \approx \rho\rho_0/(\rho_0-\rho)$. Same data as in Figure 2a.

Figure 7: $T_g$ computed from $E(\rho_g)/T_g$ =cst and the Tait equation in PDE at variable pressure versus the experimental $T_g$ values [14,17,18]: calculated $T_g$ values for : $E_1(\rho) \approx \rho^n$ : (———); $E_2(\rho) \approx \rho/b - 1$ : (--------), $E_3(\rho) \approx \rho\rho_0/(\rho_0-\rho)$ : (-.-.-.-), experimental $T_g$: (……...).

Figure 8: Comparison between the OPT α-relaxation time measured by light scattering [33] and the prediction of Eq. (6):

a) $n = 4$ and $\tilde{B} = 0, 0.5, 1.0, 1.5, 2.5, 3.48, 4.0, 5.0, 6.0$ (from top to bottom).

b) $\tilde{B} = 3.48$ and $n = 0, 1, 2, 3, 4, 5, 6, 7.5$ (from top to bottom).

Figure 9: Representation of the dielectric α-relaxation time [32] (o o o) with Eq. (6) (———) and VFT law(-------).
a) BMMPC, b) BMPC, c) KDE, d) PDE.

Figure 10: Stickel representation of the same data as in Fig. 9: a) BMMPC, b) BMPC, c) KDE, d) PDE.

| Liquid | $T_g$ (K) | $dT_g/dP$ (K MPa$^{-1}$) | $v_0$ (cm$^3$ g$^{-1}$) | $v_1$ (cm$^3$ g$^{-1}$ K$^{-1}$) | $v_2$ (cm$^3$ g$^{-1}$ K$^{-2}$) | $b_0$ (MPa) | $b_1$ (K$^{-1}$) | $\rho_g$ (g cm$^{-3}$) | $d\rho/dT$ (g cm$^{-3}$ K$^{-1}$) | $d^2\rho/dT^2$ (g cm$^{-3}$ K$^{-2}$) |
|---|---|---|---|---|---|---|---|---|---|---|
| BMMPC | 262$^a$ | 0.27$^{b,c}$ | 0.690$^a$ | 5.5 10$^{-4}$ | 4.5 10$^{-7}$ | 235 | 4.96 10$^{-3}$ | 1.114 | -6.7 10$^{-4}$ | -3,10 10$^{-7}$ |
| BMPC | 241$^a$ | 0.24$^a$ | 0.716$^a$ | 7.2 10$^{-4}$ | 0 | 201 | 4.31 10$^{-3}$ | 1.129 | -9.2 10$^{-4}$ | 1,49 10$^{-6}$ |
| KDE | 312$^f$ 315$^e$ | 0.30$^{e,f}$ | 0.895$^e$ | 4.5 10$^{-4}$ | 5.5 10$^{-7}$ | 331 | 4.46 10$^{-3}$ | 1.091 | -5.9 10$^{-4}$ | -6,61 10$^{-7}$ |
| PDE | 295$^g$ | 0.25$^g$ | 0.718$^h$ | 4.1 10$^{-4}$ | 5.1 10$^{-7}$ | 274 | 4.37 10$^{-3}$ | 1.375 | -8.2 10$^{-4}$ | -9,45 10$^{-7}$ |
| OTP | 246$^i$ | 0.26$^i$ | 0.911 | 6.4 10$^{-4}$ | 5.5 10$^{-7}$ | 189 | 4.56 10$^{-3}$ | 1.118 | -7.6 10$^{-4}$ | -3,31 10$^{-7}$ |
| DGEBA | 334$^j$ | 0.244$^j$ | 0.685$^j$ | 3.5 10$^{-4}$ | 6.8 10$^{-7}$ | 331 | 4.46 10$^{-3}$ | 1.411 | -8.6 10$^{-4}$ | -1,65 10$^{-6}$ |
| Glycerol | 187$^k$ | 0.04$^l$ | 0.793$^l$ | 3.9 10$^{-4}$ | 1.1 10$^{-7}$ | 297 | 1.92 10$^{-3}$ | 1.320 | -6.5 10$^{-4}$ | |

Table 1

| Liquid | $n$ | $b$ (g cm$^{-3}$) | $\rho_0$ (g cm$^{-3}$) | Density range (g cm$^{-3}$) | $\tilde{B}$ | Log $\tau_o$ | $m$ (calc.) | $m$ (exp.) |
|---|---|---|---|---|---|---|---|---|
| BMMPC | 7.5 | 0.97 | 1.275 | 1,09-1,13 | 1.37 | -11.4 | 69 | 59[a] |
| BMPC | 6.4 | 0.94 | 1.35 | 1,10-1,14 | 1.03 | -14.8 | 77 | 90[b] |
| KDE | 4.8 | 0.87 | 1.4 | 1,04-1,14 | 2.14 | -9.6 | 66 | 72.5[c] |
| PDE | 4.4 | 1.07 | 1.8 | 1,34-1,41 | 2.98 | -9.2 | 79 | 85[d] |
| OTP | 4 | 0.85 | 1.5 | 1,06-1,19 | 3.48 | -11.1 | 98 | 86[e] |
| DGEBA | 3.6 | 1 | 2.0 | 1,36-1,47 | - | - | - | 90[f] |
| Glycerol | 1.8 | 0.625 | 3.15* | 1,22-1,61 | 2.1 | -12.5 | 49 | 54[g] |

*this unphysical value of $\rho_0$ is due to the fact that glycerol is mainly controlled by hydrogen bonding and not by free volume effects ( see conclusion of Part II)

Table 2

| Parameters | BMMPC | BMPC | KDE | PDE |
|---|---|---|---|---|
| Eq. (6) | | | | |
| $T_g$ (K) | 261.8 | 241.5 | 314.3 | 293.7 |
| log $\tau_0$ | -11.4 | -14.8 | -9.6 | -9.2 |
| $\tilde{B}$ | 1.37 | 1.03 | 2.14 | 2.98 |
| m.s.d. | 0.022 | 0.011 | 0.0012 | 0.014 |
| Vogel-Fulcher-Tamman | | | | |
| $T_0$ (K) | 204 | 179 | 261 | 258 |
| log $\tau_{VFT}$ | -14.70 | -18.58 | -13.36 | -12.19 |
| $A$ (K) | 970.37 | 1283.19 | 827.94 | 528.89 |
| m.s.d. | 0.049 | 0.017 | 0.047 | 0.032 |

Table 3

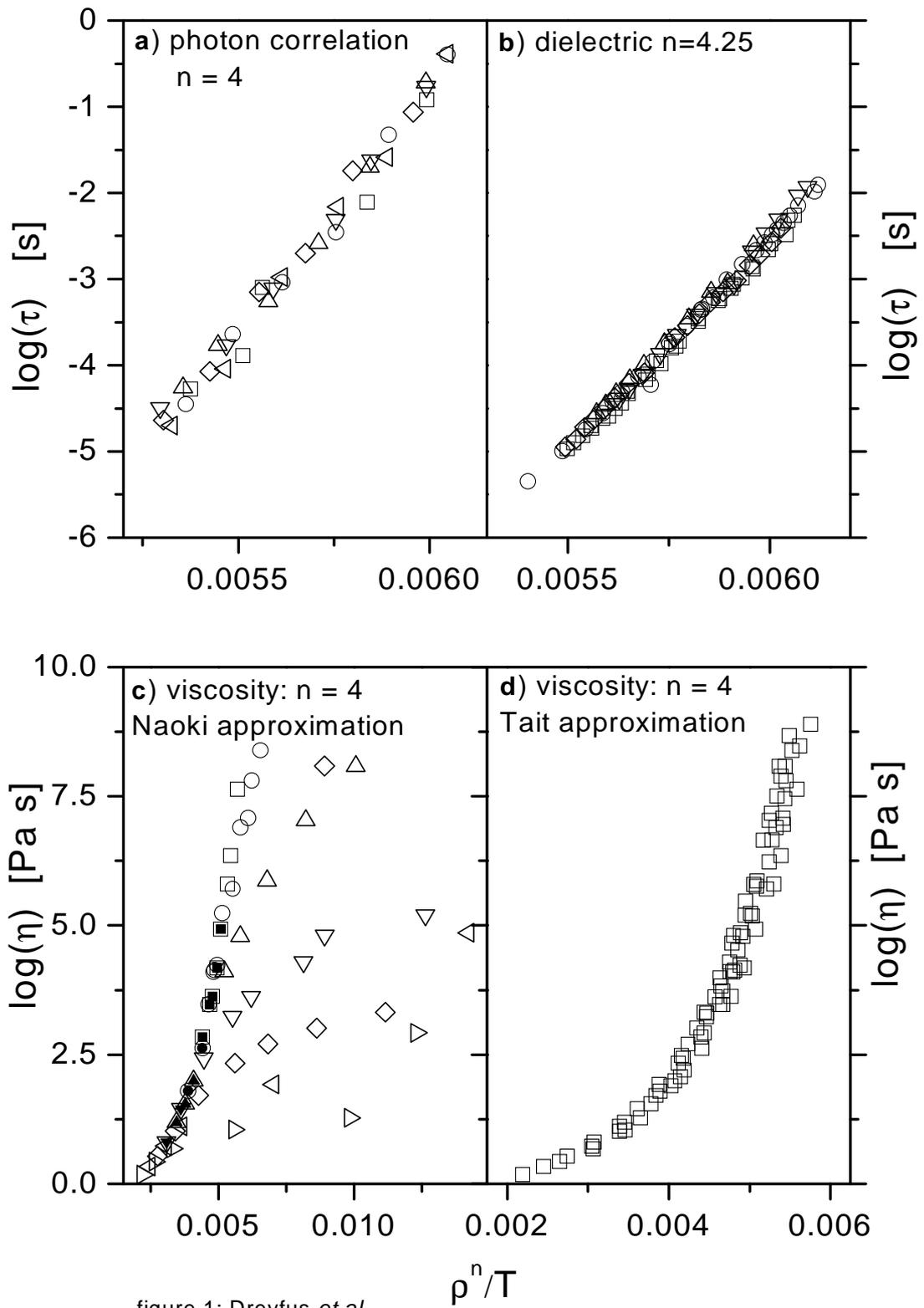

figure 1; Dreyfus *et al.*

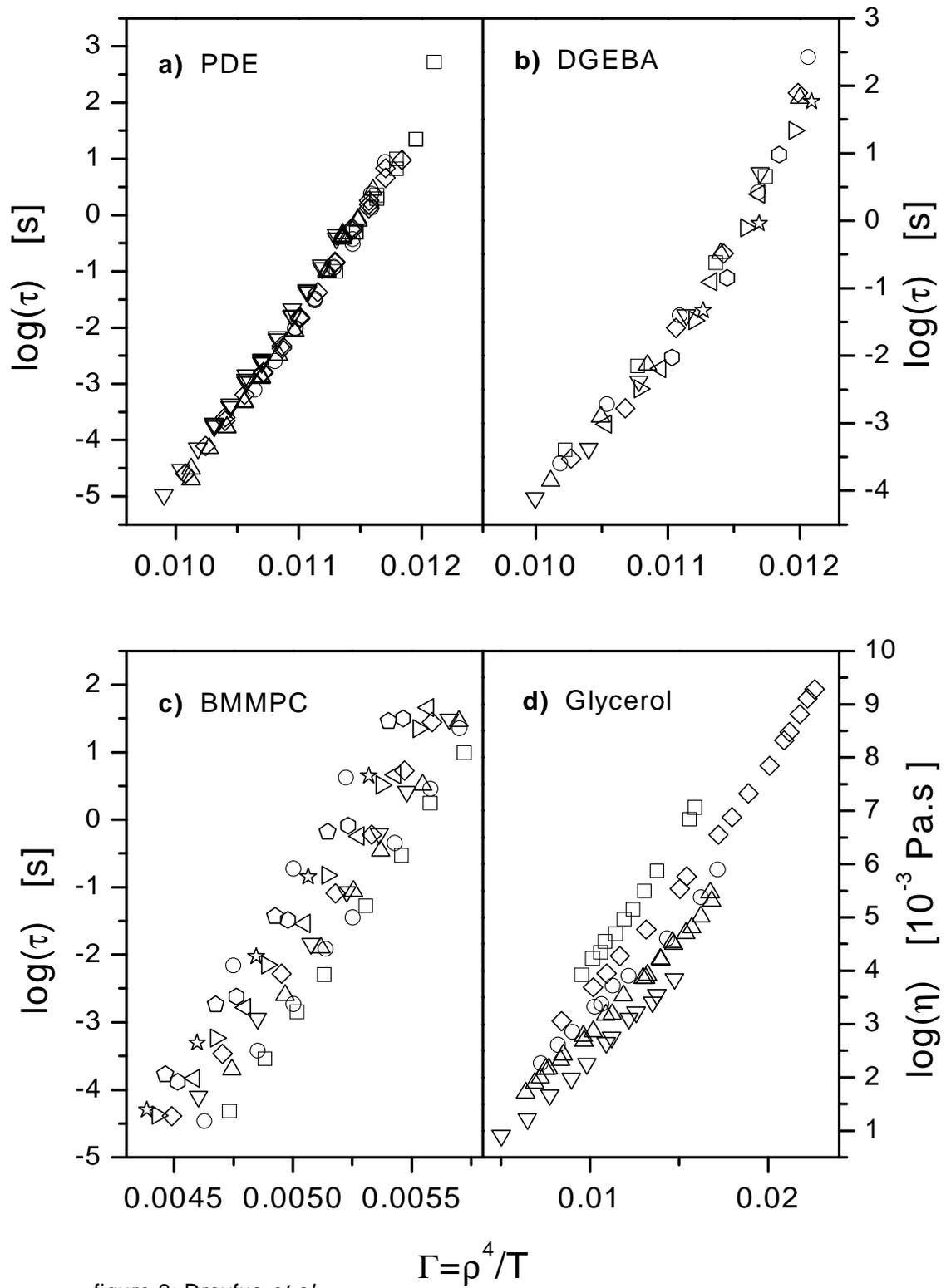

figure 2; Dreyfus *et al.*

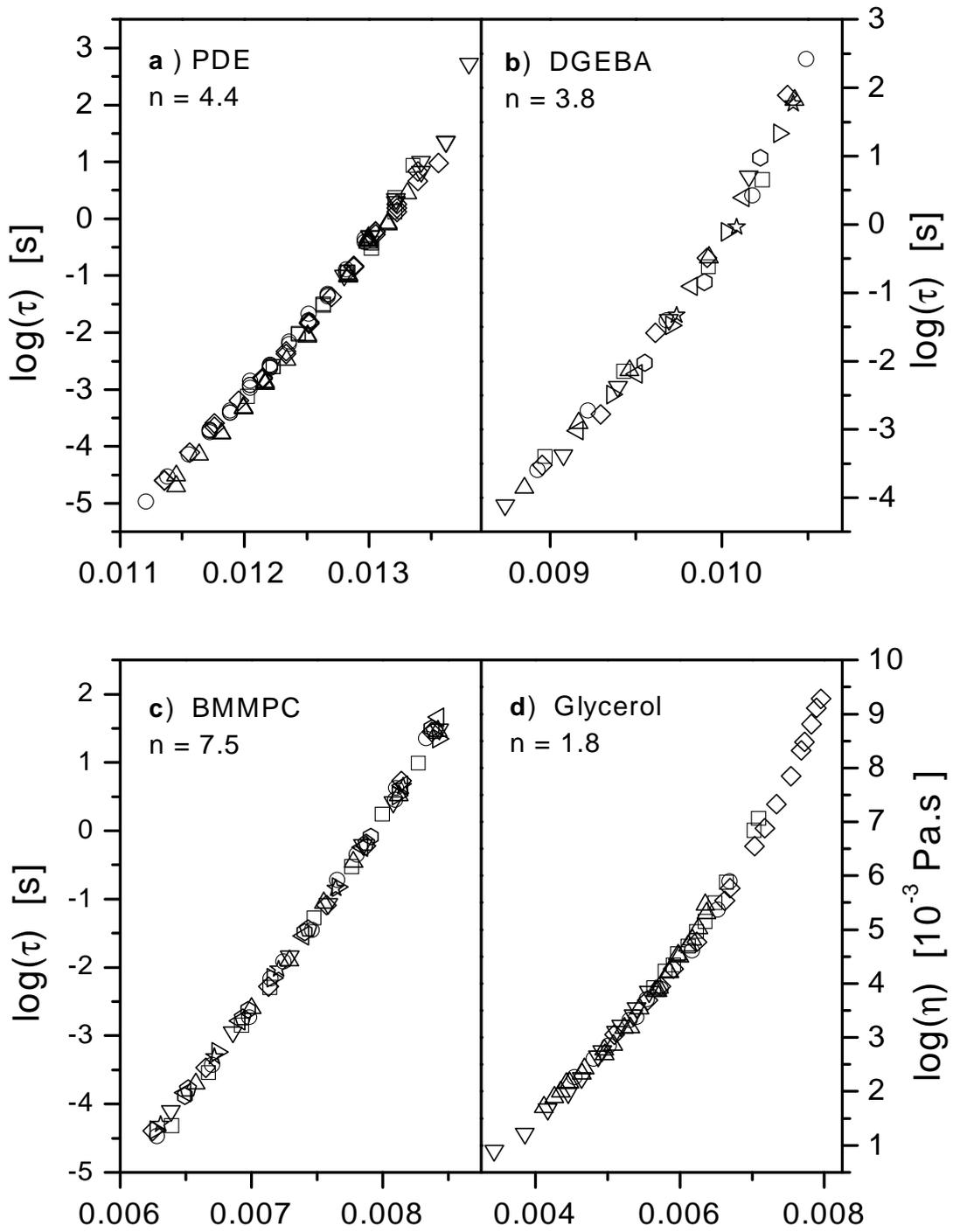

figure 3, Dreyfus *et al.*      $\Gamma = \rho^n/T$

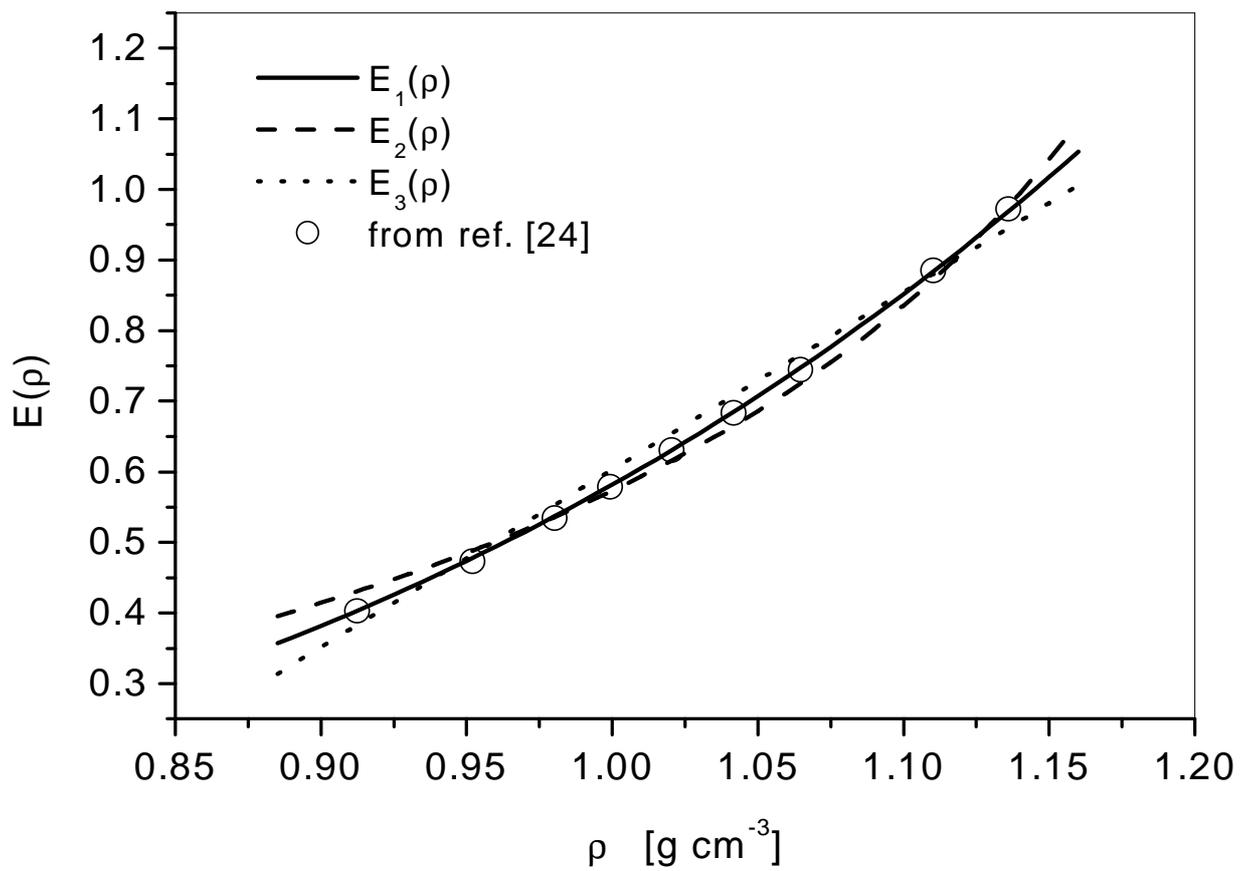

Figure 4  C.Dreyfus *et al.*

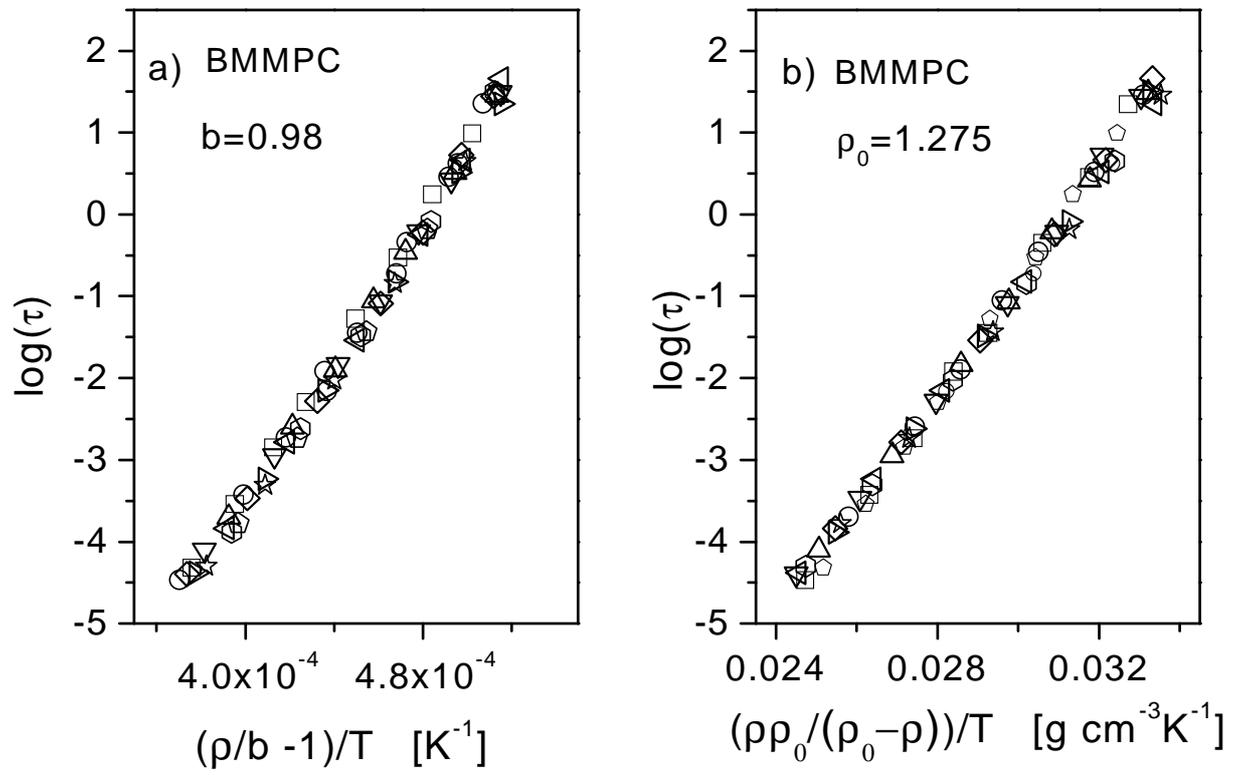

Figure 5; Dreyfus *et al.*

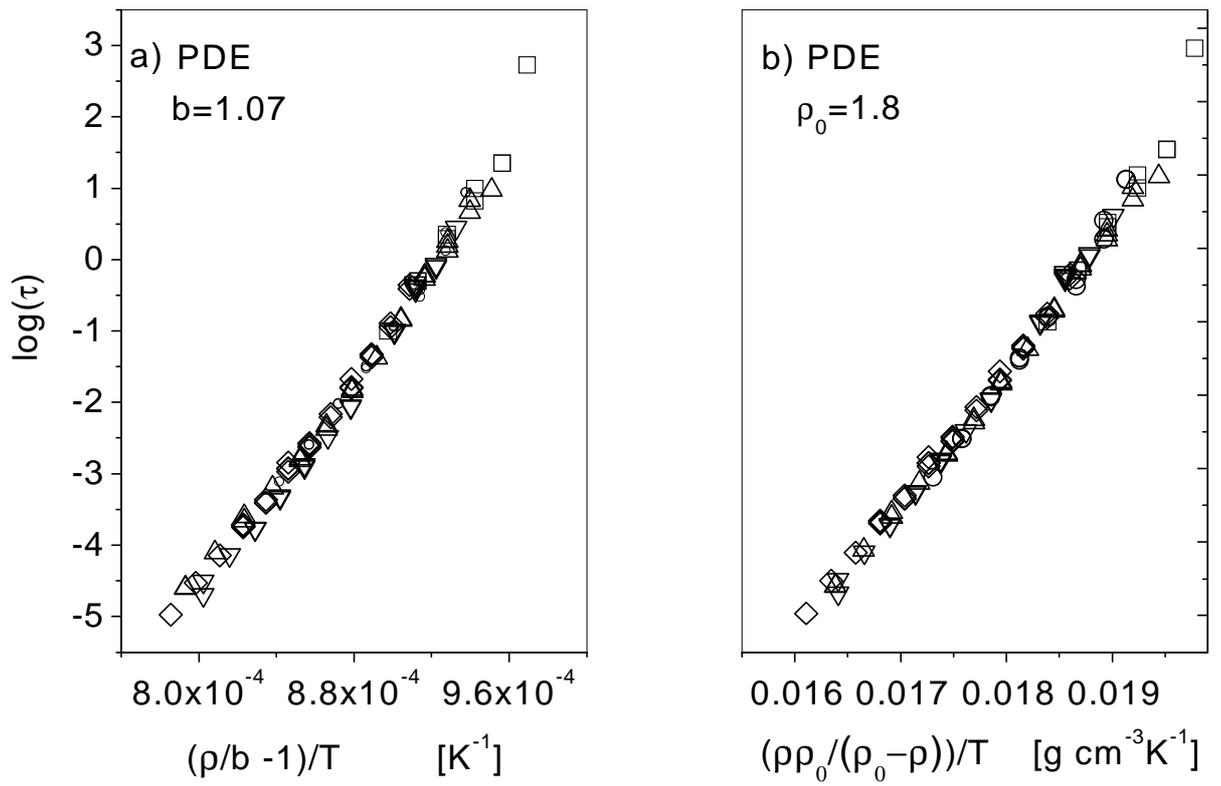

Figure 6; Dreyfus *et al.*

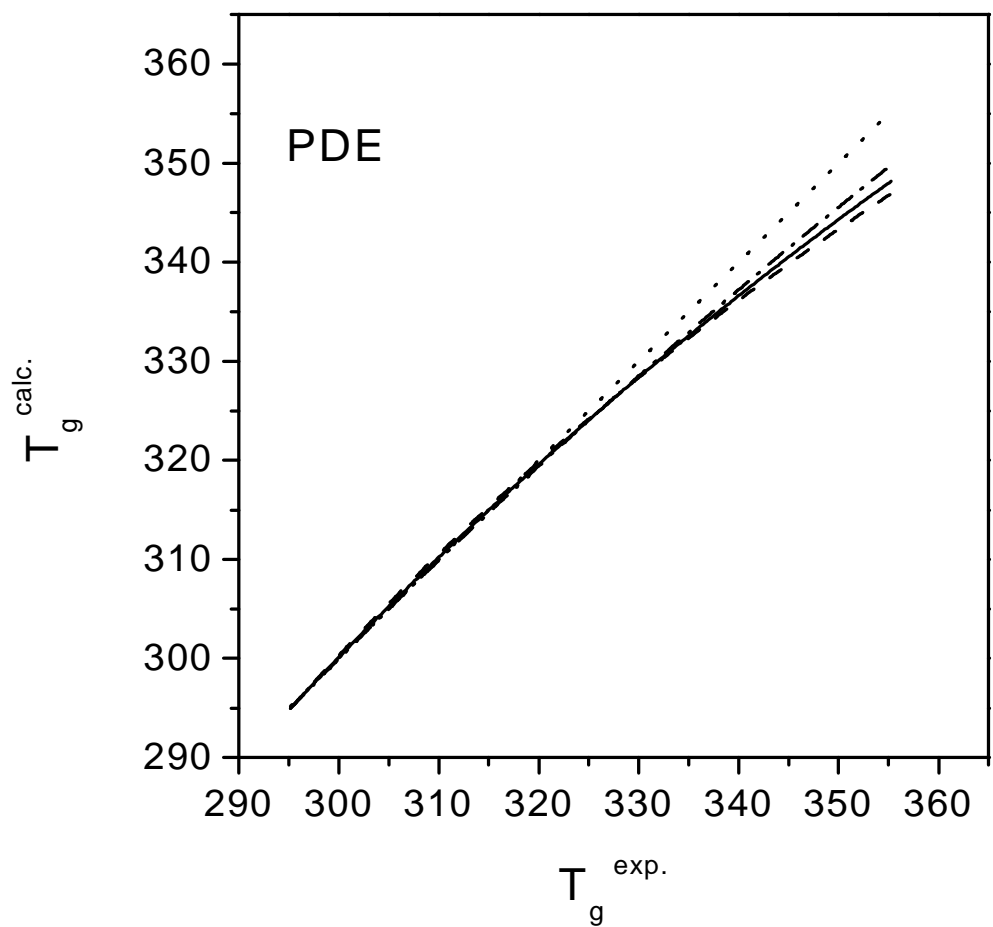

figure 7, Dreyfus *et.al.*

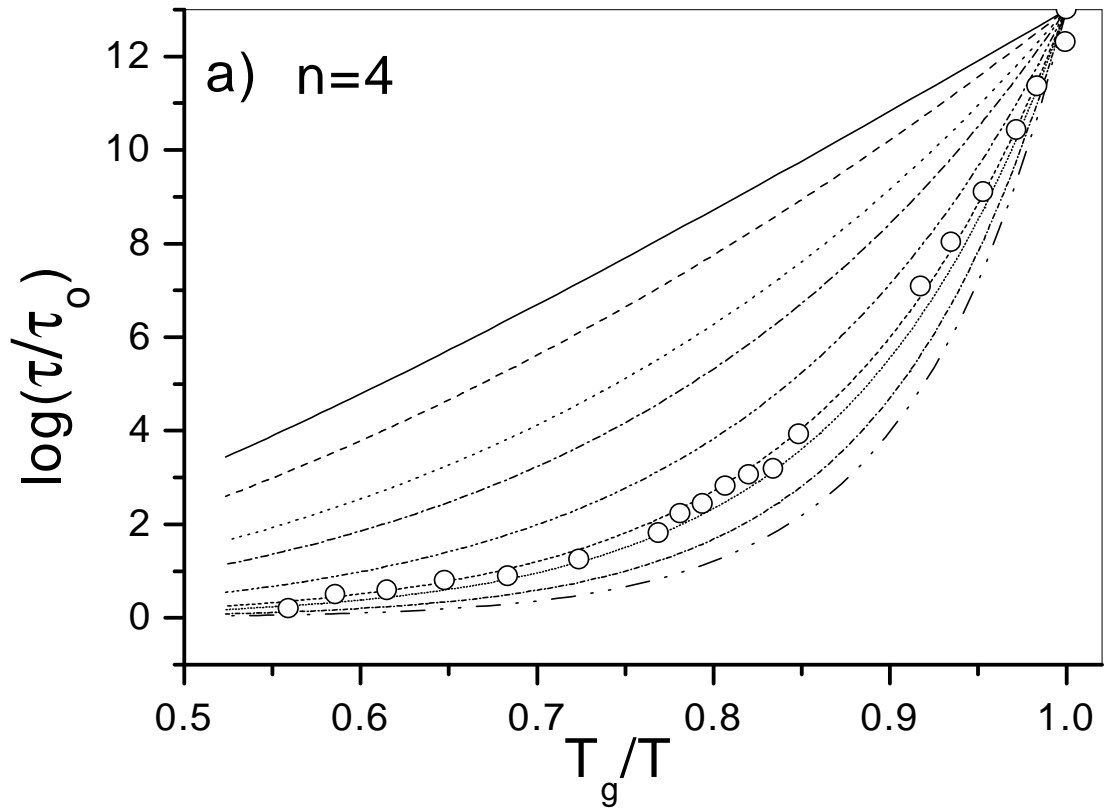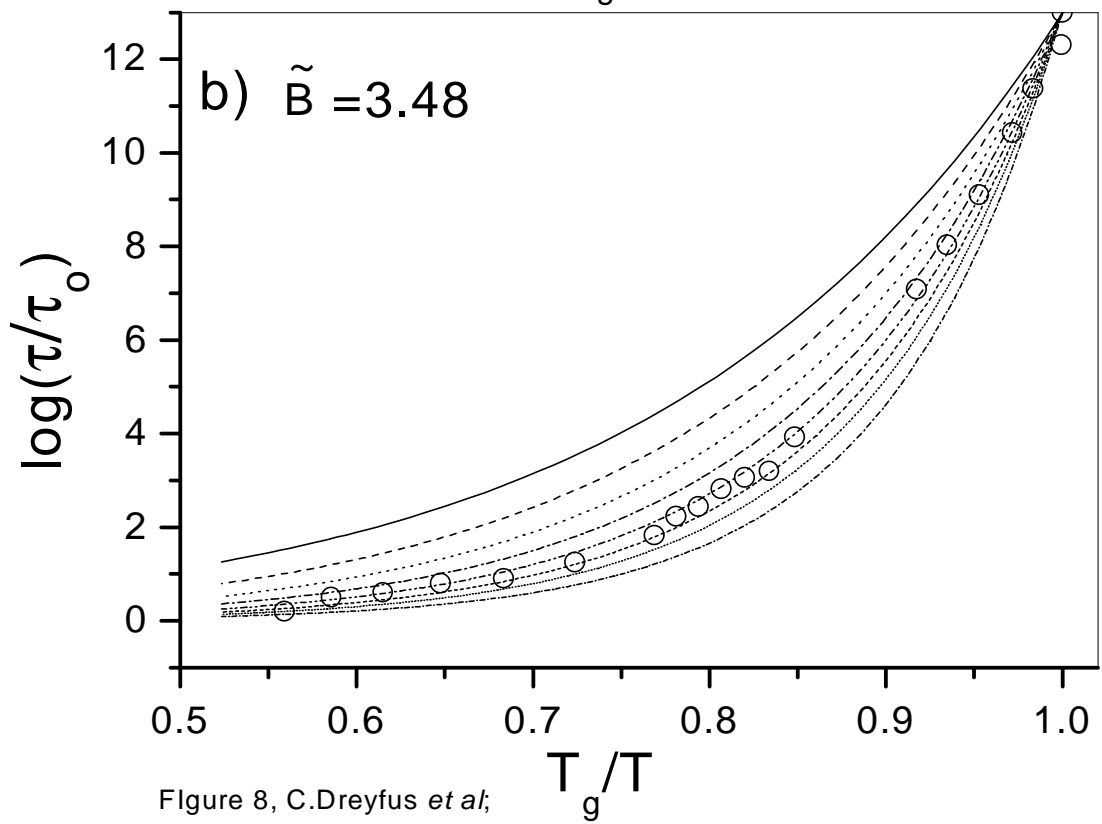

Figure 8, C.Dreyfus *et al*;

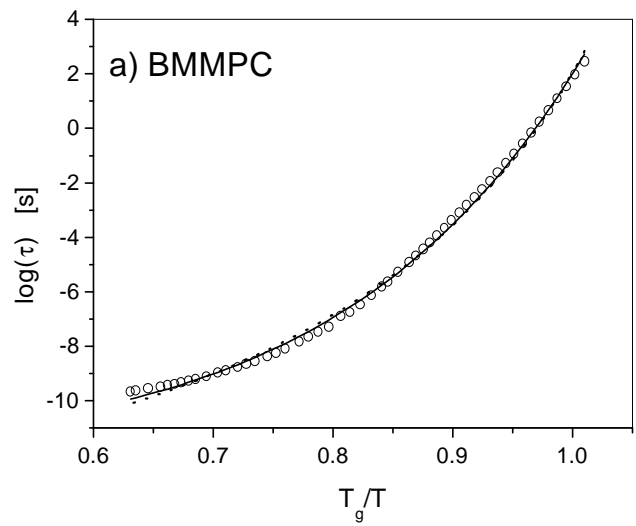
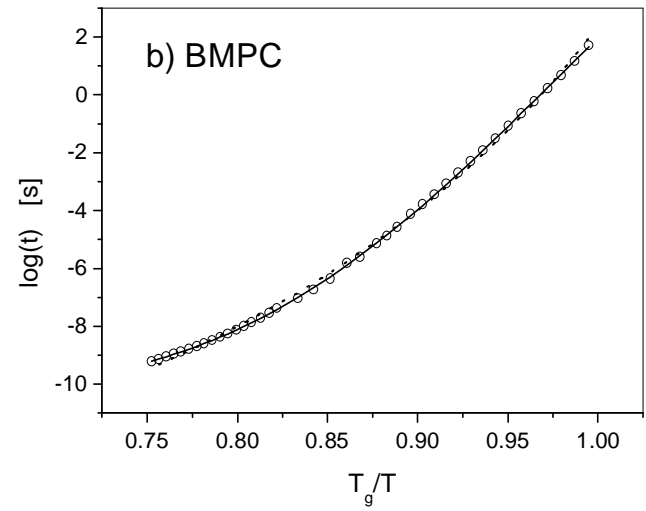
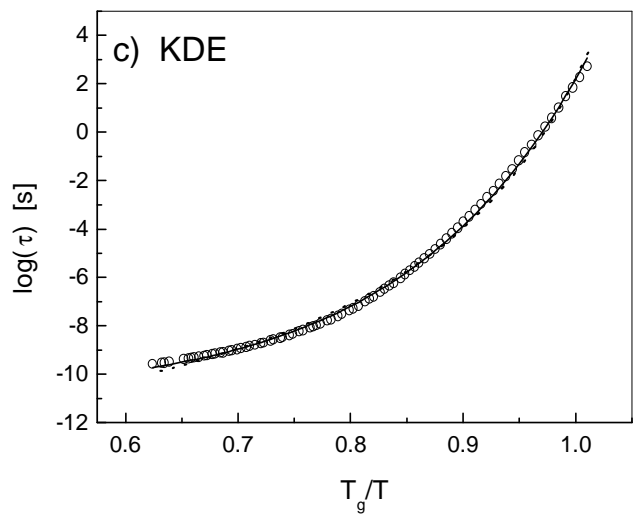
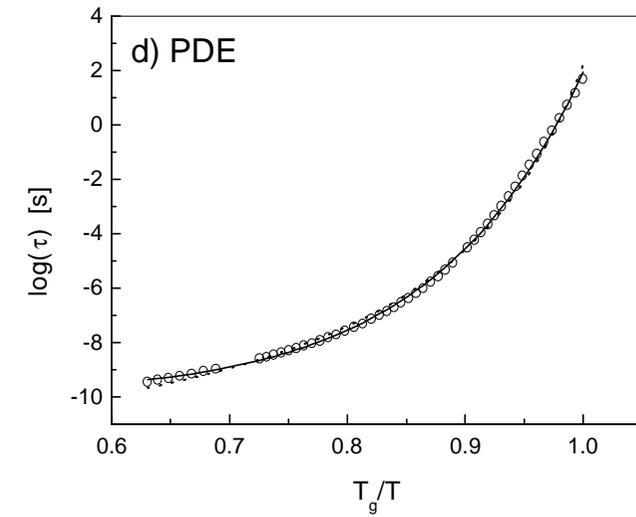

Figure 9 , C.Dreyfus *et al*;

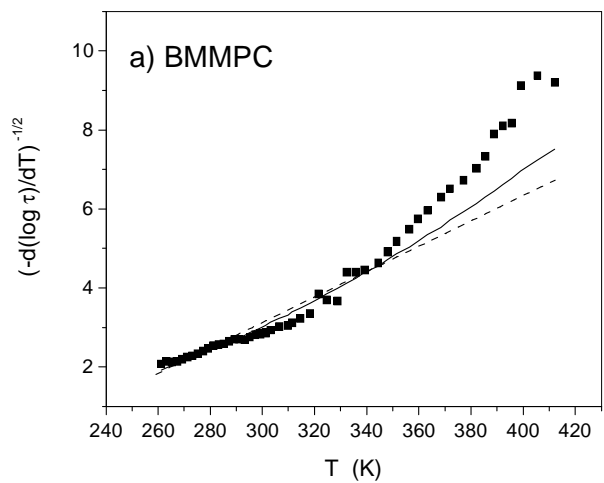
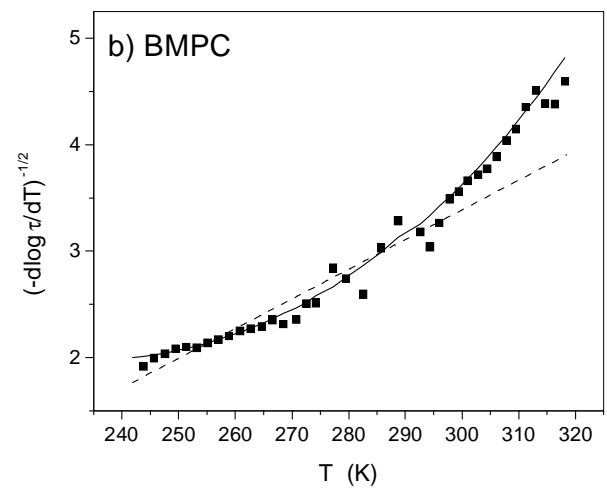
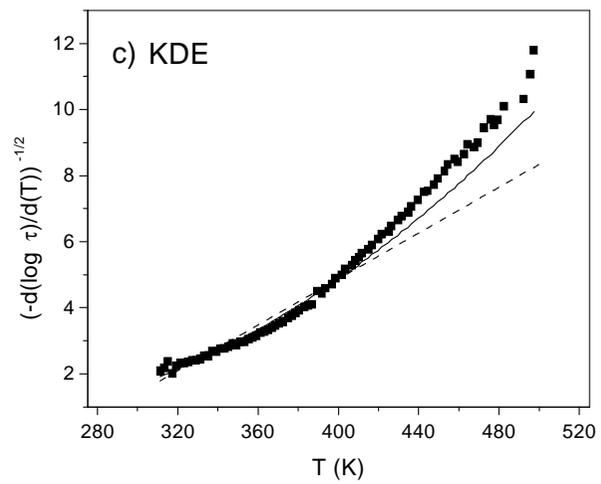
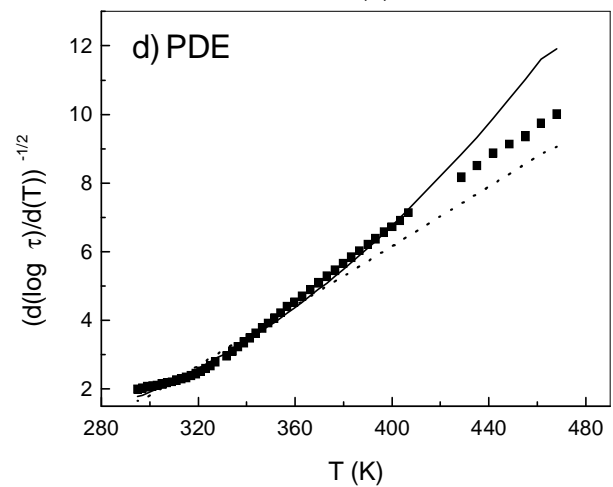

Figure 10 ; C.Dreyfus *et al.*